\documentclass[a4paper,UKenglish]{lipics-v2016}
 
\usepackage{microtype}

\usepackage{amssymb,amsmath}
\usepackage{graphicx}
\usepackage{color}
\usepackage{colortbl}
\usepackage{bm}
\usepackage{framed}
\usepackage{enumitem}
\usepackage{float}
\usepackage{hyperref}
\usepackage[english,algoruled,longend,ruled,vlined,linesnumbered]{algorithm2e}

\usepackage{lineno,setspace}
\usepackage{eqparbox}

\newsavebox{\mycodeA}
\newsavebox{\mycodeB}

\title{Fast and Simple Jumbled Indexing for Binary Run-Length Encoded Strings\footnote{This work was partially supported by CAPES/MathAmSud 021/14 and CNPq.}}
\titlerunning{Fast and Simple Jumbled Indexing for Binary RLE Strings}

\author[1]{Lu\'is Cunha}
\author[2]{Simone Dantas}
\author[3]{Travis Gagie}
\author[4]{Roland Wittler}
\author[2]{Luis Kowada}
\author[2,4]{Jens Stoye}

\affil[1]{Universidade Federal do Rio de Janeiro, Rio de Janeiro, Brazil\\
  \texttt{lfignacio@cos.ufrj.br}}
\affil[2]{Universidade Federal Fluminense, Niter\'oi, Brazil\\
  \texttt{sdantas@im.uff.br,luis@ic.uff.br}}
\affil[3]{Universidad Diego Portales and CEBIB, Santiago, Chile\\
  \texttt{travis.gagie@mail.udp.cl}}
\affil[4]{Universit\"at Bielefeld, Bielefeld, Germany\\
  \texttt{\{roland.wittler,jens.stoye\}@uni-bielefeld.de}
}
\authorrunning{Lu\'is Cunha \emph{et al.}} 

\Copyright{Lu\'is Cunha \emph{et al.}}


\EventEditors{Tomasz Kociumaka, Jakub Radoszewski, Wojciech Rytter and Tomasz Wale\'n}
\EventNoEds{4}
\EventLongTitle{28th Annual Symposium on Combinatorial Pattern Matching (CPM 2017)}
\EventShortTitle{CPM 2017}
\EventAcronym{CPM}
\EventYear{2017}
\EventDate{July 4--6, 2017}
\EventLocation{Warsaw, Poland}
\EventLogo{}
\SeriesVolume{}
\ArticleNo{}

\begin{document}

\maketitle

\begin{abstract}
Important papers have appeared recently on the problem of indexing binary strings for jumbled pattern matching, and further lowering the time bounds in terms of the input size would now be a breakthrough with broad implications.  We can still make progress on the problem, however, by considering other natural parameters.
Badkobeh et al.\ (IPL, 2013) and Amir et al.\ (TCS, 2016) gave algorithms that index a binary string in $O (n + \rho^2 \log \rho)$ time, where $n$ is the length and $\rho$ is the number of runs, and Giaquinta and Grabowski (IPL, 2013) gave one that runs in $O (n + \rho^2)$ time.
In this paper we propose a new and very simple algorithm that also runs in $O(n + \rho^2)$ time and can be extended either so that the index returns the position of a match (if there is one), or so that the algorithm uses only $O (n)$ bits of space.
\end{abstract}

\section{Introduction}

Since its introduction at the 2009 Prague Stringology Conference~\cite{burcsi2012approximate,cicalese2009searching}, the problem of indexed binary jumbled pattern matching has been discussed in many top conferences and journals.  It asks us to preprocess a binary string such that later, given a number of 0s and a number of 1s, we can quickly report whether there exists a substring with those numbers of 0s and 1s and, optionally, return the position of one such substring or possibly even all of them.  The na\"ive preprocessing algorithm takes quadratic time but researchers have reduced that bound to $O (n^2 / \log n)$~\cite{burcsi2010table,moosa2010indexing}, $O (n^2 / \log^2 n)$~\cite{moosa2012sub}, \(O(n^2 / 2^{\Omega (\sqrt{\log n / \log \log n})})\)~\cite{bremner2014necklaces,hermelin2014binary} and finally $O (n^{1.859})$ with randomization or $O (n^{1.864})$ without~\cite{chan2015clustered}.

Researchers have also looked at indexing for approximate matching~\cite{cicalese2013indexes,cicalese2012near}, indexed jumbled pattern matching over larger alphabets~\cite{amir2014hardness,kociumaka2013efficient}, indexing labelled trees and other structures~\cite{cicalese2013indexes,durocher2014indexed,gagie2015binary}, and how to index faster when the (binary) input string is compressible.  Gagie et al.~\cite{gagie2015binary} gave an algorithm that runs in $O (g^{2 / 3} n^{4 / 3})$ when the input is represented as a straight-line program with $g$ rules, and Badkobeh et al.~\cite{badkobeh2013binary} gave one that runs in $O (n + \rho^2 \log \rho)$ time when the input consists of $\rho$ runs, i.e., maximal unary substrings (we will denote later as $\rho$ the number of maximal substrings of 1s, for convenience).  Giaquinta and Grabowski~\cite{giaquinta2013new} gave two algorithms: one runs in $O (\rho^2 \log k + n / k)$ time, where $k$ is a parameter, and produces an index that uses $O (n / k)$ extra space and answers queries in $O (\log k)$ time; the other runs in $O (n^2 \log^2 (w) / w)$ time, where $w$ is the size of a machine word.  Amir et al.~\cite{amir2016algorithms} gave an algorithm that runs in $O (\rho^2 \log \rho)$ time when the input is a run-length encoded binary string, or $O (n + \rho^2 \log \rho)$ time when it is a plain binary string; it builds an index that takes $O (\rho^2)$ words and answers queries in $O (\log \rho)$ time, however.  Very recently, Sugimoto et al.~\cite{sugimoto2017computing} considered the related problems of finding Abelian squares, Abelian periods and longest common Abelian factors, also on run-length encoded strings.

We first review some preliminary notions in Section~\ref{sec:pre}.  We present our main result in Section~\ref{sec:alg}: a new and very simple indexing algorithm that runs in $O (n + \rho^2)$ time, which matches Giaquinta and Grabowski's algorithm with the parameter \(k = 1\) and is thus tied as the fastest known when \(\rho = \Omega (n^{0.5}) \cap o (n^{0.932})\) and the smallest straight-line program for the input has \(\omega (\rho^3 / n^2)\) rules.  For an input string of up to ten million bits, for example, if the average run-length is three or more then \(\rho < n^{0.932}\).  Our algorithm takes only 17 lines of pseudocode, making it a promising starting point for investigating other possible algorithmic features.  In Section~\ref{sec:witness}, for example, we show how to extend our algorithm to store information that lets us report the position of a match (if there is one).  Finally, in Section~\ref{sec:workspace}, we show how we can alternatively adapt it to use only $O (n)$ bits of space.

\section{Preliminaries}\label{sec:pre}

Consider a string \(s \in \{0, 1\}^n\).  We denote by \(s [i \cdots j]\) the substring of $s$ consisting of the $i$th through $j$th characters, for \(1 \leq i \leq j \leq n\); if \(i = j\), we can also write simply \(s [i]\).  Cicalese et al.~\cite{burcsi2012approximate,cicalese2009searching} observed that, if we slide a window of length $k$ over $s$, the number of 1s in the window can change by at most 1 at each step.  It follows that if \(s [i \cdots i + k - 1]\) contains $x$ copies of 1 and \(s [j \cdots j + k - 1]\) contains $z$ copies of 1 with \(i \leq j\) then, for $y$ between $x$ and $z$ (notice $x$ could be smaller than, larger than, or equal to $z$), there is a substring of length $k$ in \(s [i \cdots j + k - 1]\) with exactly $y$ copies of 1.  This immediately implies the following theorem:

\begin{theorem}
\label{thm:table}
Let $x$ and $z$ be the minimum and maximum numbers of 1s in any substring of length $k$. There is a substring of length $k$ with $y$ copies of 1 if and only if $x \leq y \leq z$.
\end{theorem}

By Theorem~\ref{thm:table}, if we compute and store, for \(1 \leq k \leq n\), the minimum and maximum numbers of 1s in a substring of $s$ of length $k$ then later, given a number of 0s and a number of 1s, we can report in constant time whether there exists a substring with that many 0s and 1s.  For example, if \(s = 010101110011\) then, as $k$ goes from 1 to \(n = 12\), the minimum and maximum numbers of 1s are \(0, 0, 1, 2, 2, 3, 4, 4, 5, 5, 6, 7\) and \(1, 2, 3, 3, 4, 4, 5, 5, 6, 6, 7, 7\), respectively. Since the fifth numbers in these lists are 2 and 4, we know, there are substrings of length 5 with exactly 2, 3 and 4 copies of 1, but none with 0, 1 or 5 (or more than 5, obviously).

Cicalese et al.~\cite{cicalese2013indexes} noted that, if we also store the positions of the substrings with the minimum and maximum numbers of 1s and a bitvector for $s$ that supports constant time rank queries, then via binary search in \(O (\log n)\) time we can find an example of a substring with any desired numbers of 0s and 1s, called a {\em witness} if such a substring exists. (The query \(\mathrm{rank} (i)\) returns the number of 1s in \(s [1 \cdots i]\); see, e.g.,~\cite{navarro2016compact} for more details of rank queries on bitvectors.) For example, suppose we want to find a substring of length 5 with exactly 3 copies of 1 in our example string $s$. We have stored that there are substrings of length 5 with 2 and 4 copies of 1 starting at positions 1 and 4, respectively, so we know there is a substring of length 5 with exactly 3 copies of 1 starting in \(s [1 \cdots 4]\). We choose \(\lfloor (1 + 4) / 2 \rfloor = 2\) and check how many 1s there are in \(s [2 \cdots 2 + 5 - 1 = 6]\) via two rank queries. In this case, the answer is 3, so we have found a witness in one step; otherwise, we would know there is a witness starting in \(s [3 \cdots 4]\) and we would recurse on that interval.

The same authors noted that in each step, the lists of minimum and maximum numbers can only stay the same or increment, so we can represent each list as a bitvector of length $n$ and support access to it using rank queries.  For example, the bitvector for the list of minimum numbers in our example is \(001101101011\), so \(\mathrm{rank} (i)\) returns the $i$th number in the list.  Since an $n$-bit bitvector takes $O (n)$ bits of space, it follows that we can store our index in $O (n)$ bits and still support constant-time queries, if we do not want a witness.  We note, however, that even though the input $s$ takes $n$ bits and the resulting index takes $O (n)$ bits, all previous constructions have used $\Omega (n)$ {\em words} in the worst case.

A \emph{run} in $s$ is a maximal unary substring and the run-length encoding $\mathrm{rle} (s)$ is obtained by replacing each run by a copy of the character it contains and its length. Although $\rho$ is usually used to denote the number of runs, for convenience, we use it to denote only the number of runs of 1s --- about half its normal value for binary strings --- and consider $s$ to begin and end with (possibly empty) runs of 0s. For example, for our example string the run-length encoding is $0^1 1^1 0^1 1^1 0^1 1^3 0^2 1^2 0^0$ and \(\rho = 4\) (instead of 9). We denote the lengths of the runs of zeroes and 1s as \(z [0], \ldots, z [\rho]\) and \(o [1], \ldots, o [\rho]\), respectively.

\section{Basic Indexing}\label{sec:alg}

Since finding substrings with the minimum numbers of 1s is symmetric to finding substrings with the maximum numbers of 1s (e.g., by taking the complement of the string), we describe how, given a binary run-length encoded string \(s [1 \cdots n]\), we can build a table \(T [1 \cdots n]\) such that \(T [k] = f (k)\), where \(f (k)\) denotes the maximum number of 1s in a substring of $s$ of length $k$.

The complete pseudo-code of our algorithm --- only 17 lines --- is shown as Algorithm~\ref{alg:index}.  The starting point of our explanation and proof of correctness is the observation that, if the bit immediately to the left of a substring is a 1, we can shift the substring one bit left without decreasing the number of 1s; if the first bit of the substring is a 0, then we can shift the substring one bit right (shortening it on the right if necessary) without decreasing the number of 1s.  It follows that, for \(1 \leq k \leq n\), there is a substring of length at most $k$ containing \(f (k)\) copies of 1 and starting at the beginning of a run of 1s.  Since we can remove any trailing 0s from such a substring also without changing the number of 1s, there is such a substring that also ends in a run of 1s. Therefore we have the following lemma:

\begin{algorithm}[t]
  \DontPrintSemicolon
  \For{$i=1, \ldots, n$}{
    $T[i] = 0$\;
  }
  \For{$i=1, \ldots, \rho$}{
    $ones = o[i]$\;
    $zeros = 0$\;
    $T[ones] = ones$\;
    \For{$j=i+1, \ldots, \rho$}{
      $ones\; +\!= o[j]$\;
      $zeros\; +\!= z[j-1]$\;
      \If{$ones > T[ones+zeros]$}{
        $T[ones+zeros] = ones$\;
      }
    }
  }
  \For{$i=n-1, \ldots, 1$}{
    \If{$T[i] < T[i+1]-1$}{
       $T[i] = T[i+1]-1$\;
    }
  }
  \For{$i=2, \ldots, n$}{
    \If{$T[i] < T[i-1]$}{
      $T[i] = T[i-1]$
    }
  }
\caption{\label{alg:index}Building the index table $T$ of string $s$.}
\end{algorithm}

\begin{lemma}
\label{lem:starting}
For \(1 \leq k \leq n\), there is a substring of length at most $k$ containing \(f (k)\) copies of 1, starting at the beginning of a run of 1s and ending in a run of 1s.
\end{lemma}

Applying Lemma~\ref{lem:starting} immediately yields an $O (n \rho)$-time algorithm: set \(T [1 \cdots n]\) to all 0s; for each position $i$ at the beginning of a run of 1s and each position $j$ in a run of 1s, set \(T [j - i + 1] = \max (T [j - 1 + 1], s [i] + \cdots + s [j])\); finally, because $f$ is non-decreasing, make a pass over $T$ from \(T [2]\) to \(T [n]\) setting each \(T [i] = \max (T [i], T [i - 1])\).  Computing the number \(s [i] + \cdots + s [j]\)  of 1s in a substring \(s [i \dots j]\) starting at the beginning of a run of 1s and ending in a run of 1s is easy to do from the run-length encoding in amortized constant time.

To speed this preliminary algorithm up to run in $O (n + \rho^2)$ time, we first observe that, if $\ell$ is the length of a substring starting at the beginning of a run of 1s, ending in a run of 1s and containing \(f (\ell)\) copies of 1, and \(d > \ell\) is the length of a substring starting at the beginning of a run of 1s and ending at the end of a run of 1s, then \(f (\ell) \geq f (d) - d + \ell\). (In fact this is true for any $\ell$ and \(d \geq \ell\), simply because \(f (x + 1) \leq f (x) + 1\) for all $x$.)  We then observe that, for some such $d$, we have \(f (\ell) = f (d) - d + \ell\).  To see why, consider any substring \(s [i \cdots j]\) of length $\ell$ starting at the beginning of a run of 1s, ending within a run of 1s and containing \(f (\ell)\) copies of 1: let $d$ be the length of the substring starting at \(s [i]\) and ending at the end of the run of 1s containing \(s [i + \ell - 1]\), so \(f (\ell) = f (d) - d + \ell\).

\begin{lemma}
\label{lem:ending}
If $\ell$ is the length of a substring starting at the beginning of a run of 1s, ending in a run of 1s and containing \(f (\ell)\) copies of 1, and \(d > \ell\) is the length of a substring starting at the beginning of a run of 1s and ending at the end of a run of 1s, then \(f (\ell) \geq f (d) - d + \ell\).  Furthermore, for some such $d$, we have \(f (\ell) = f (d) - d + \ell\).
\end{lemma}

With Lemma~\ref{lem:ending}, we can compute the number \(s[i] + \cdots + s[j]\)  of 1s in each substring \(s [i \dots j]\) starting at the beginning of a run of 1s and ending in a run of 1s, in a total of \(O (n + \rho^2)\) time: again, set \(T [1 \cdots n]\) to all 0s; for each position $i$ at the beginning of a run of 1s and each position $j$ at the end of a run of 1s, set \(T [j - i + 1] = \max (T [j - 1 + 1], s [i] + \cdots + s [j])\); make a pass over $T$ from \(T [n - 1]\) to \(T [1]\) setting each \(T [i] = \max (T [i], T [i + 1] - 1)\).  Computing the number \(s [i] + \cdots + s [j]\) of 1s in a substring \(s [i \cdots j]\) starting at the beginning of a run of 1s and ending at the end of a run of 1s is again easy to do from the run-length encoding in amortized constant time.

Combining Lemmas~\ref{lem:starting} and~\ref{lem:ending}, we have a complete algorithm for computing $T$ in $O (n + \rho^2)$ time: set \(T [1 \cdots n]\) to all 0s; for each position $i$ at the beginning of a run of 1s and each position $j$ at the end of a run of 1s, set \(T [j - i + 1] = \max (T [j - 1 + 1], s [i] + \cdots + s [j])\); make a pass over $T$ from \(T [n - 1]\) to \(T [1]\) setting each \(T [i] = \max (T [i], T [i + 1] - 1)\) (which sets \(T [\ell]\) correctly for every length $\ell$ of a substring starting at the beginning of a run of 1s, ending in a run of 1s and containing \(f (\ell)\) copies of 1); and make a pass over $T$ from \(T [2]\) to \(T [n]\) setting each \(T [i] = \max (T [i], T [i - 1])\) (which sets every entry in $T$ correctly).  Once we have $T$, we can convert it into a bitvector in $O (n)$ time.  Summarizing our results so far, we have the following theorem, which we adapt in later sections:

\begin{theorem}
\label{thm:main}
Given a binary string $s$ of length $n$ containing $\rho$ runs of 1s, we can build an $O (n)$-bit index for constant-time jumbled pattern matching in $O (n + \rho^2)$ time.
\end{theorem}

Now we examine how our algorithm works on our example \(s = 010101110011\). First we set all entries of $T$ to 0, then we loop through the runs of 1s and, for each, loop through the runs of 1s not earlier, computing distance from the start of the first to the end of the second and the number of 1s between those positions. While doing this, we set \(T [1] = 1\), the number of 1s from the start to the end of the first run of 1s; \(T [3] = 2\), the number of 1s from the start of the first run of 1s to the end of the second run of 1s; \(T [7] = 5\), the number of 1s from the start of the first run of 1s to the end of the third run of 1s; \(T [11] = 7\), the number of 1s from the start of the first run of 1s to the end of the fourth run of 1s; \(T [5] = 4\), the number of 1s from the start of the second run of 1s to the end of the third run; etc.  When we have finished this stage, \(T = [1, 2, 3, 0, 4, 0, 5, 0, 6, 0, 7, 0]\).  We then make a pass over $T$ from right to left, setting each \(T [i] = \max (T [i], T [i + 1] - 1)\).  After this stage, \(T = [1, 2, 3, 3, 4, 4, 5, 5, 6, 6, 7, 0]\).  Finally, we make a pass over $T$ from left to right, setting each \(T [i] = \max (T [i], T [i - 1])\). This fills in \(T [12]\) and leaves $T$ correctly computed as \(T = [1, 2, 3, 3, 4, 4, 5, 5, 6, 6, 7, 7]\).

\section{Witnessing Index}
\label{sec:witness}

As described in Section~\ref{sec:pre}, if together with computing the minimum and maximum number of 1s in a substring of length $k$ for \(1 \leq k \leq n\), we also store the positions of substrings of length~$k$ with those numbers of 1s, and a single bitvector for $s$, then, together with confirming that $s$ contains a substring with a given number of 0s and 1s (if it does), we can give the starting position of such a substring, still in constant time.

In this section, we show how to modify our algorithm from Section~\ref{sec:alg} to build also a table \(P [1..n]\) such that \(P [k]\) is the starting position of a substring of length $k$ containing \(f (k)\) copies of 1s.  Computing and storing the starting position of a substring of length $k$ with the minimum number of 1s is symmetric.

First, notice that during the first stage of Algorithm~\ref{alg:index}, whenever we set \(T [k] = f (k)\), we have found a substring of length $k$ containing \(f (k)\) copies of 1, so we can set \(P [k]\) at the same time.  Now consider the second stage of the algorithm, in which we make a right-to-left pass over $T$ setting \(T [i] = \max (T [i], T [i + 1] - 1)\) for \(1 \leq i \leq n - 1\).  When we start this stage, for every positive entry in $T$ we have set the corresponding entry in $P$.  Therefore, by induction, whenever we set \(T [i] = T [i + 1] - 1\), we have \(P [i + 1]\) set to the starting position of a substring of length \(i + 1\) containing \(T [i + 1]\) copies of 1.  The substring of length $i$ starting at \(P [i + 1]\) contains at least \(T [i + 1] - 1\) copies of 1, so we can set \(P [i] = P [i + 1]\).  In the last stage of the algorithm, in which we make a left-to-right pass over $T$, we can almost use the same kind of argument and simply copy $P$ values when we copy $T$ values, except that we must ensure the starting positions we copy are far enough to the left of the end of the string (i.e., that the substrings have the correct lengths). Our modified algorithm is shown as Algorithm~\ref{alg:witness} --- still only 25 lines --- and we now have the following theorem:

\begin{theorem}
\label{thm:witness}
Given a binary string $s$ of length $n$ containing $\rho$ runs of 1s, we can build an $O (n)$-word index for constant-time jumbled pattern matching {\em with witnessing} in $O (n + \rho^2)$ time.
\end{theorem}

Running our modified algorithm on our example \(s = 010101110011\), in the first stage we set \(T = [1, 2, 3, 0, 4, 0, 5, 0, 6, 0, 7, 0]\) and, simultaneously, \(P = [2, 11, 6, 0, 4, 0, 2, 0, 4, 0, 2, 0]\), where 0 indicates an unset value in $P$.  In the second stage, we set \(T = [1, 2, 3, 3, 4, 4, 5, 5, 6, 6, 7, 0]\) and \(P = [2, 11, 6, 4, 4, 2, 2, 4, 4, 2, 2, 0]\).  Finally, in the third stage, we fill in \(T [12] = T [11]\), but we cannot just set \(P [12] = P [11] = 2\) because \(s [2 \cdots n = 12]\) has length only 11, so we set \(P [12] = 1\).

\begin{algorithm}[t]
  \DontPrintSemicolon
  \For{$i=1, \ldots, n$}{
    $T[i] = 0$\;
  }
  $p = z[0]$\;
  \For{$i=1, \ldots, \rho$}{
    $ones = o[i]$\;
    $zeros = 0$\;
    $T[ones] = ones$\;
    $P[ones] = p$\;
    \For{$j=i+1, \ldots, \rho$}{
      $ones\; +\!= o[j]$\;
      $zeros\; +\!= z[j-1]$\;
      \If{$ones > T[ones+zeros]$}{
        $T[ones+zeros] = ones$\;
        $P[ones+zeros] = p$\;
      }
    }
  $p\; +\!= ones[i]+zeros[i]$\;
  }
  \For{$i=n-1, \ldots, 1$}{
    \If{$T[i] < T[i+1]-1$}{
       $T[i] = T[i+1]-1$\;
       $P[i] = P[i+1]$\;
    }
  }
  \For{$i=2, \ldots, n$}{
    \If{$T[i] < T[i-1]$}{
      $T[i] = T[i-1]$\;
      $P[i] = P[i-1]$\;
      \If{$P[i]+i > n$}{
        $P[i] = n-i$\;
      } 
    }
  }
\caption{\label{alg:witness}Building the tables $T$ and $P$ for $s$.}
\end{algorithm}

\section{Reducing Workspace}
\label{sec:workspace}

It is frustrating that both $s$ and the index described in Theorem~\ref{thm:main} take $O (n)$ bits, but we use $O (n)$ words to build the index.  In this section, we show how to reduce this workspace to $O (n)$ bits also, without increasing the time bound for construction by more than a constant factor.

Suppose we divide $T$ into blocks of size \(\lg (n) / 2\) and modify our algorithm such that, whenever we set a value \(T [i]\), we ensure that each value \(T [j]\) in the same block with \(j < i\) is at least \(T [i] - i + j\) and each value \(T [j]\) in the same block with \(i < j\) is at least \(T [i]\).  Since we would eventually set each such \(T [j]\) to a value at least as great during the normal execution of the algorithm, this does not change its correctness, apart from perhaps slowing it down by an $O (\log n)$ factor.

For any two consecutive values \(T [i]\) and \(T [i + 1]\) in the same block now, however, we have \(T [i] \leq T [i + 1] \leq T [i] + 1\). We can thus store each block by storing its first value and a binary string of length \(\lg (n) / 2\) whose bits indicate where the values in the block increase.  Therefore, we need a total of only $O (n)$ bits to store all the blocks.

Notice that, if we increase a value \(T [i]\) by more than \(\lg (n) / 2\), we reset the first value \(T [h]\) of the block to be \(T [i] - i + h\), set the leading bits of the block to 1s to indicate that the values increase until reaching \(T [i]\), and set the later bits of the block to 0s to indicate that the values remain equal to \(T [i]\) until the end of the block.  Therefore, we can speed the algorithm up to run in $O (n + \rho^2)$ time again, by using a universal table of size \(2^{\lg (n) / 2} \log^{O (1)} n = o (n^{1 / 2 + \epsilon})\) to decide how to update blocks when we set values in them.

\begin{theorem}
\label{thm:workspace}
Given a binary string $s$ of length $n$ containing $\rho$ runs of 1s, we can build an $O (n)$-bit index for constant-time jumbled pattern matching in $O (n + \rho^2)$ time using $O (n)$ bits of workspace.
\end{theorem}

In fact, it seems possible to make the algorithm run in $O (n + \rho^2)$ time and $O (n)$ bits of space even without a universal table, using AC0 operations on words that are available on standard architectures.

This workspace reduction makes little sense for a string as small as our example \(s = 010101110011\) but, for the sake of argument, suppose we partition our array $T$ for it into three blocks of length 4 each.  We keep \(T [1]\), \(T [5]\) and \(T [9]\) stored explicitly and represent the other entries of $T$ implicitly with three 3-bit binary strings \(B_1\), \(B_2\) and \(B_3\).  Initially we set \(T [1] = T [5] = T [9] = 0\) and \(B_1 = B_2 = B_3 = 000\).  Recall from Section~\ref{sec:alg} that we first set \(T [1] = 1\), the number of 1s from the start to the end of the first run of 1s.  At this point, we do not need to change \(B_1\).  We then set \(T [3] = 2\) --- the number of 1s from the start of the first run of 1s to the end of the second run of 1s --- by setting \(B_1 = 010\): starting from \(T [1] = 1\), this encodes \(T [2] = T [1] + 0 = 1\), \(T [3] = T [1] + 0 + 1 = 2\) and \(T [4] = T [1] + 0 + 1 + 0 = 2\).  Next we set \(T [7] = 5\) --- the number of 1s from the start of the first run of 1s to the end of the third run of 1s --- by setting \(T [5] = 3\) and \(B_2 = 110\): starting from \(T [5] = 3\), this encodes \(T [6] = T [5] + 1 = 4\), \(T [7] = T [5] + 1 + 1 = 5\) and \(T [8] = T [5] + 1 + 1 + 0 = 5\).  Continuing like this, we set \(T [11] = 7\) by setting \(T [9] = 5\) and \(B_3 = 110\); set \(T [5] = 4\) and \(B_2 = 010\); etc.  When we are finished this stage, \(T [1] = 1\), \(T [5] = 4\) and \(T [9] = 6\), and \(B_1 = 110\), \(B_2 = 010\) and \(B_3 = 010\), encoding \(T = [1, 2, 3, 3, 4, 4, 5, 5, 6, 6, 7, 7]\).  In this case the final right-to-left and left-to-right passes have no effect, but there are cases (e.g., when we do not set any values in a certain block) when they are still necessary.

\end{document}